\documentclass[aps,pra,reprint,superscriptaddress,nofootinbib,floatfix]{revtex4-2}
\usepackage{amsmath,amssymb,amsfonts,bm}
\usepackage{graphicx}
\usepackage{hyperref}
\usepackage{physics}
\usepackage{braket}
\usepackage{xcolor}
\hypersetup{colorlinks=true,linkcolor=blue,citecolor=blue,urlcolor=blue}

\newcommand{\Css}{C_{\mathrm{ss}}}
\newcommand{\ii}{\mathrm{i}}
\newcommand{\ee}{\mathrm{e}}
\newcommand{\numax}{\nu_{\mathrm{max}}}
\newcommand{\varmax}{\varphi_{\mathrm{max}}}
\newcommand{\HN}{Hatano--Nelson}
\newcommand{\trn}{\tilde n}
\newcommand{\trnu}{\tilde \nu}
\newcommand{\trA}{\tilde{\mathcal A}}

\begin{document}

\title{Natural-orbital locking reveals hidden steady-state skin order in Gaussian open fermion chains}
\author{Y. T. Wang}
\author{X. Z. Zhang}
\email{zhangxz@tjnu.edu.cn}
\affiliation{College of Physics and Materials Science, Tianjin Normal University, Tianjin 300387, China}

\begin{abstract}
Nonreciprocal relaxation matrices can have skin-localized right eigenmodes, but their imprint on a mixed steady state is not fixed by the density profile alone. We develop an exact steady-state theory for number-conserving Gaussian fermion chains and show that the dominant natural orbital of the correlation matrix provides a mode-resolved diagnostic of hidden skin order. The steady-state correlator admits a biorthogonal decomposition in terms of the left and right eigenmodes of the relaxation matrix $X$ and the source matrix $Y$. This formula separates three ingredients: slow rapidity denominators, source loading by left eigenmodes, and real-space geometry from right eigenmodes. For a local pump, the pump position is read by the left modes, whereas the selected profile is drawn by the right modes. In a single-slow-mode regime, the dominant natural orbital locks to the Euclidean-normalized slow right mode. The density can follow the same boundary trend, but it is a less selective incoherent sum over occupied natural orbitals. We verify this selection law in a nonreciprocal \HN{} chain and show that, in a nonreciprocal SSH chain, the selected natural orbital crosses over from a topological edge candidate to a slow bulk-skin candidate. These results identify natural-orbital locking as a steady-state diagnostic of nonreciprocal localization in Gaussian open fermion chains.
\end{abstract}

\maketitle

\setcounter{secnumdepth}{0}  
\paragraph{Introduction.}
The non-Hermitian skin effect has changed the way boundary sensitivity is understood in nonreciprocal lattice systems~\cite{HatanoNelsonPRL1996,HatanoNelsonPRB1997,YaoWang2018,Gong2018,BergholtzRMP2021}. In a Lindblad setting, however, the long-time object is a density matrix rather than a wave function. For quadratic fermionic Lindbladians the steady state is Gaussian, and its one-body content is encoded in a correlation matrix~\cite{Prosen2008,Prosen2010,Song2019,Haga2021,Yang2022,McDonald2022}. This distinction raises a basic question. If the relaxation matrix has skin-localized right eigenmodes, which steady-state observable most directly identifies the selected localization pattern?
\begin{figure}[t]
\centering
\includegraphics[width=0.5\textwidth]{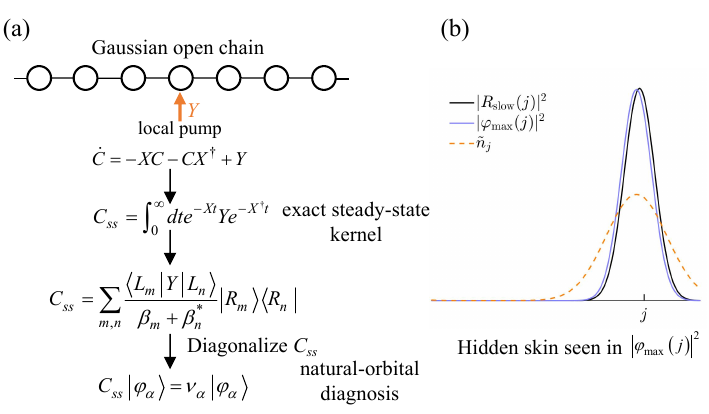}
\caption{Conceptual definition of hidden steady-state skin order. A finite Gaussian open chain is driven by a local pump $Y=\Gamma\ket{s}\bra{s}$ and relaxes under a non-Hermitian matrix $X$. The exact steady-state kernel can be written as the integral in Eq.~\eqref{eq:Cintegral} or as the biorthogonal expansion in Eq.~\eqref{eq:Css_biorth}. Diagonalizing $\Css$ produces natural orbitals. The schematic comparison on the right illustrates the diagnostic used below. The Euclidean-normalized slow right mode $|\widehat R_{\rm slow}(j)|^2$ and the dominant natural-orbital profile $|\varphi_{\max}(j)|^2$ identify the selected skin profile, while the normalized density $\trn_j$ is a less selective diagonal contraction of the same correlator.}
\label{fig:schematic}
\end{figure}
The local density is a natural observable, but it is only a diagonal contraction of the steady-state correlator. We write
\begin{equation}
 C_{ij}^{\mathrm{ss}}=\Tr\!\left(\rho_{\mathrm{ss}}c_j^\dagger c_i\right),
 \qquad n_j^{\mathrm{ss}}=(\Css)_{jj} .
\end{equation}
The full correlator has its own eigenmodes,
\begin{equation}
 \Css\ket{\varphi_\alpha}=\nu_\alpha\ket{\varphi_\alpha},
 \label{eq:natural_orbitals}
\end{equation}
which are the natural orbitals of the Gaussian steady state. For a physical fermionic Gaussian state $0\leq \nu_\alpha\leq 1$. The density is reconstructed as
\begin{equation}
 n_j^{\mathrm{ss}}=\sum_\alpha \nu_\alpha |\varphi_\alpha(j)|^2 .
 \label{eq:density_from_orbitals}
\end{equation}
Thus the density is not a mode-resolved object. It is an incoherent sum over all occupied natural orbitals. When one occupation dominates, the density may look similar to the leading orbital. When subleading orbitals carry appreciable weight, the density becomes smoother. In either case, the dominant natural orbital is the sharper object for asking which steady-state mode has been selected.

In this Letter we derive an exact selection theory for this dominant orbital. The steady-state correlator can be expanded in the biorthogonal eigenbasis of the relaxation matrix $X$. The expansion separates the steady-state kernel into slow denominators, left-mode loading, and right-mode geometry. This separation is the key physical point. A local source is sampled by the left eigenmodes, while the spatial profile of a selected steady-state component is inherited from the right eigenmodes. In a single-slow-mode regime, this structure forces the dominant natural orbital of $\Css$ to lock to the Euclidean-normalized slow right eigenmode.

We demonstrate the mechanism in two complementary examples. The nonreciprocal \HN{} chain gives the minimal bulk-skin setting and allows a direct comparison between the exact source dependence and the analytical loading law. The nonreciprocal SSH chain adds dimerization and edge sectors, so that the same natural-orbital diagnostic distinguishes an edge-locking regime from a bulk-skin-locking regime. Figure~\ref{fig:schematic} summarizes the logic. The normalized density
\begin{equation}
 \trn_j=\frac{n_j^{\mathrm{ss}}}{\sum_\ell n_\ell^{\mathrm{ss}}}
 \label{eq:norm_density}
\end{equation}
is shown as a density-level summary of the same mixed state, not as the defining diagnostic. Explicit microscopic jump constructions for the target families are given in the Supplemental Material.

\setcounter{secnumdepth}{0}  
\paragraph{Exact steady-state theory.}
\begin{figure*}[t]
\centering
\includegraphics[width=0.78\textwidth]{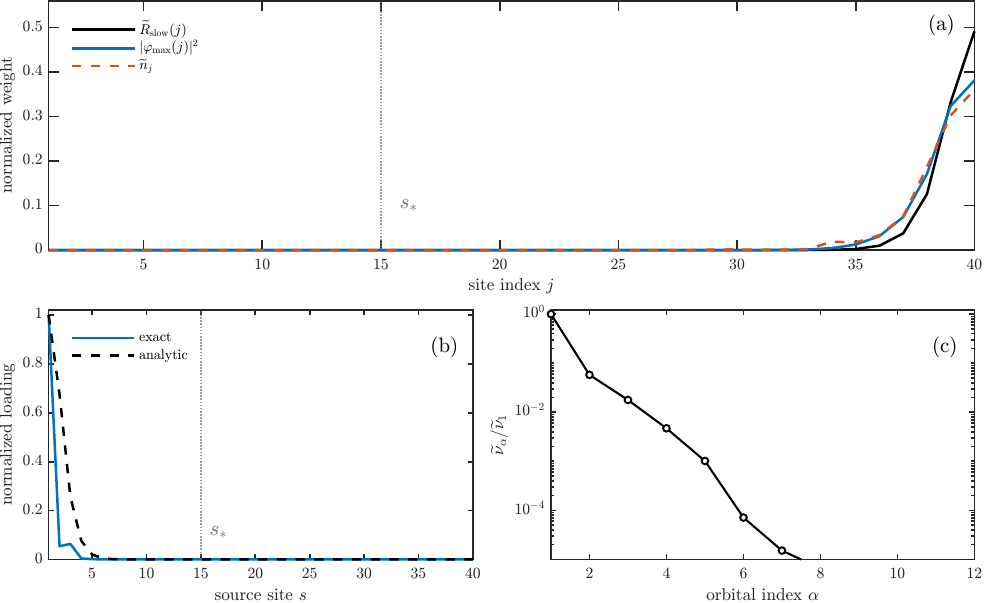}
\caption{\HN{} chain and validation of the source-selection law. The relaxation matrix is Eq.~\eqref{eq:XHN}, and all panels use $N=40$, $t_R=1$, $t_L=0.17$, $\kappa=0.91$, $\Gamma=0.03$, $s_\ast=15$. (a) Euclidean-normalized slow right-mode profile $|\widehat R_{\rm slow}(j)|^2$, dominant natural-orbital profile $|\varphi_{\max}(j)|^2$, and normalized density $\trn_j$ for a pump at $s_\ast=15$. The natural orbital tracks the selected right mode directly, while the density is the diagonal contraction of the same steady-state mixture. (b) Source-position scan comparing the normalized exact loading $\nu_{\max}(s)/\max_{s'}\nu_{\max}(s')$ with the normalized analytic prediction $\mathcal A_1(s)/\max_{s'}\mathcal A_1(s')$. The source dependence is fixed by the left-mode weight $|L_1(s)|^2$. (c) Normalized occupation spectrum $\trnu_\alpha=\nu_\alpha/\nu_{\max}$, showing the separation of a dominant natural orbital.}
\label{fig:HN}
\end{figure*}
We consider a number-conserving quadratic Lindblad problem with Hamiltonian
\begin{equation}
 H=\sum_{i,j=1}^{N}h_{ij}c_i^\dagger c_j,
 \qquad h=h^\dagger,
\end{equation}
and linear gain and loss jumps. The one-body correlator
\begin{equation}
 C_{ij}(t)=\Tr\!\left[\rho(t)c_j^\dagger c_i\right]
\end{equation}
obeys the closed equation
\begin{equation}
 \dot C=-XC-CX^\dagger+Y .
 \label{eq:compactCeq}
\end{equation}
Here
\begin{equation}
 X=\ii h+\frac{\Gamma^-+\Gamma^+}{2},
 \qquad Y=\Gamma^+,
 \label{eq:XandY}
\end{equation}
where $\Gamma^-$ and $\Gamma^+$ are the loss and gain Gram matrices. Equation~\eqref{eq:compactCeq} is derived in the Supplemental Material from the microscopic Lindblad equation.

The steady state satisfies
\begin{equation}
 X\Css+\Css X^\dagger=Y .
\end{equation}
If all eigenvalues of $X$ have positive real parts, the solution is
\begin{equation}
 \Css=\int_0^\infty dt\,\ee^{-Xt}Y\ee^{-X^\dagger t} .
 \label{eq:Cintegral}
\end{equation}
Let
\begin{equation}
 X\ket{R_n}=\beta_n\ket{R_n},
 \qquad
 X^\dagger\ket{L_n}=\beta_n^*\ket{L_n},
\end{equation}
with biorthogonal normalization
\begin{equation}
 \langle L_m|R_n\rangle=\delta_{mn},
 \qquad \sum_n\ket{R_n}\bra{L_n}=I .
\end{equation}
Inserting the spectral decomposition of $\ee^{-Xt}$ into Eq.~\eqref{eq:Cintegral} gives
\begin{equation}
 \Css=\sum_{m,n}
 \frac{\mel{L_m}{Y}{L_n}}{\beta_m+\beta_n^*}
 \ket{R_m}\bra{R_n} .
 \label{eq:Css_biorth}
\end{equation}
This formula is the central result. The factor $(\beta_m+\beta_n^*)^{-1}$ controls temporal accumulation, $\mel{L_m}{Y}{L_n}$ controls source loading, and $\ket{R_m}\bra{R_n}$ carries the real-space geometry.

For a local pump,
\begin{equation}
 Y=\Gamma\ket{s}\bra{s},
 \qquad \Gamma>0,
\end{equation}
Eq.~\eqref{eq:Css_biorth} becomes
\begin{equation}
 \Css=\Gamma\sum_{m,n}
 \frac{L_m^*(s)L_n(s)}{\beta_m+\beta_n^*}
 \ket{R_m}\bra{R_n} .
 \label{eq:Css_localpump}
\end{equation}
The pump position therefore enters through left eigenmodes, while the selected spatial profile is drawn by right eigenmodes. This left-right separation is absent in Hermitian steady-state intuition and is the origin of the source-selection effect below.

Suppose that one mode, labeled $0$, is parametrically slower than the rest and that $L_0(s)\neq0$. Then Eq.~\eqref{eq:Css_localpump} is dominated by the $m=n=0$ term,
\begin{equation}
 \Css\simeq \mathcal A_0(s)\ket{R_0}\bra{R_0},
 \qquad
 \mathcal A_0(s)=\frac{\Gamma |L_0(s)|^2}{2\Re\,\beta_0} .
 \label{eq:singlemodeCss}
\end{equation}
Because $\ket{R_0}$ is biorthogonally normalized and need not have unit Euclidean norm, the corresponding dominant natural orbital is
\begin{equation}
 \ket{\varmax}\simeq \ket{\widehat R_0},
 \qquad
 \ket{\widehat R_0}=\frac{\ket{R_0}}{\sqrt{\langle R_0|R_0\rangle}} .
 \label{eq:locking}
\end{equation}
The leading occupation is
\begin{equation}
 \numax\simeq \mathcal A_0(s)\langle R_0|R_0\rangle .
\end{equation}
Thus the orbital profile is fixed by the right mode, while the source dependence is fixed by the left-mode loading factor.

In the figures we compare natural orbitals with Euclidean-normalized right modes through
\begin{equation}
 \mathcal O_n=\left|\langle \widehat R_n|\varmax\rangle\right|^2 .
 \label{eq:overlap}
\end{equation}
We also use
\begin{equation}
 \trnu_\alpha=\frac{\nu_\alpha}{\numax},
 \qquad
 \mathcal A_n(s)=\frac{\Gamma |L_n(s)|^2}{2\Re\,\beta_n} .
 \label{eq:An}
\end{equation}
Equation~\eqref{eq:density_from_orbitals} then fixes the relation between the density and the natural-orbital profile. The density can be close to $|\varmax(j)|^2$ in a strongly single-orbital regime, but it remains a less selective contraction of $\Css$.

\setcounter{secnumdepth}{0}  
\paragraph{Hatano--Nelson and nonreciprocal SSH chains.}
\begin{figure*}[t]
\centering
\includegraphics[width=0.76\textwidth]{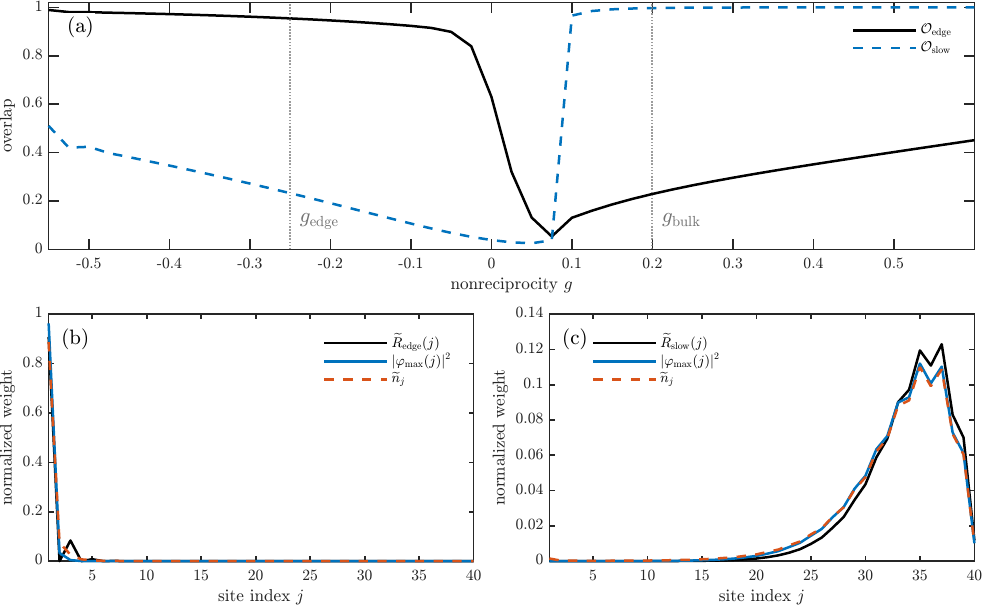}
\caption{Nonreciprocal SSH chain and crossover between edge locking and bulk-skin locking. The target matrix is Eq.~\eqref{eq:XSSH}, the common parameters are $N=20$, $t_1=0.50$, $t_2=1.00$, $\kappa=1.50$, $\Gamma=10^{-8}$, $(m,A)=(1,A)$, and the overlaps are defined in Eq.~\eqref{eq:OedgeOslow}. (a) Crossover of $\mathcal O_{\rm edge}$ and $\mathcal O_{\rm slow}$ as a function of nonreciprocity $g$. The dotted lines mark $g_{\rm edge}=-0.25$ and $g_{\rm bulk}=0.20$. (b) At $g_{\rm edge}$, the dominant natural orbital tracks the Euclidean-normalized edge candidate $|\widehat R_{\rm edge}(j)|^2$. (c) At $g_{\rm bulk}$, it tracks the Euclidean-normalized slow bulk-skin mode $|\widehat R_{\rm slow}(j)|^2$. In both panels the normalized density $\trn_j$ is shown for comparison. Panels (b) and (c) use different vertical scales.}
\label{fig:SSH}
\end{figure*}

We now apply the theory to two target relaxation matrices. The \HN{} chain isolates the bulk skin mechanism and provides a direct test of the source-selection law. The SSH chain adds dimerization and topological edge modes, turning the steady-state problem into a selection crossover between distinct candidate modes.

For the nonreciprocal \HN{} chain,
\begin{equation}
 X_{\rm HN}=\kappa I-t_R\sum_{j=1}^{N-1}\ket{j+1}\bra{j}
 -t_L\sum_{j=1}^{N-1}\ket{j}\bra{j+1},
 \label{eq:XHN}
\end{equation}
with $t_R>t_L>0$, define
\begin{equation}
 r=\sqrt{\frac{t_R}{t_L}},
 \qquad S=\sum_{j=1}^{N}r^j\ket{j}\bra{j} .
\end{equation}
Then
\begin{equation}
 S^{-1}X_{\rm HN}S=
 \kappa I-\sqrt{t_Rt_L}\sum_{j=1}^{N-1}
 \left(\ket{j+1}\bra{j}+\ket{j}\bra{j+1}\right),
\end{equation}
which is a Hermitian open chain. The reference eigenfunctions and eigenvalues are
\begin{align}
& \phi_n(j)=\sqrt{\frac{2}{N+1}}\sin\frac{n\pi j}{N+1},\\
 &\beta_n=\kappa-2\sqrt{t_Rt_L}\cos\frac{n\pi}{N+1}.
\end{align}
Transforming back gives
\begin{equation}
 R_n(j)=r^j\phi_n(j),
 \qquad L_n(j)=r^{-j}\phi_n(j) .
 \label{eq:HNRL}
\end{equation}
For $r>1$, right modes accumulate at the right edge and left modes at the left edge. With $Y=\Gamma\ket{s}\bra{s}$, the source dependence of the slowest mode is $\mathcal A_1(s)=\Gamma |L_1(s)|^2/(2\Re\,\beta_1)$. Thus the pump reads a left-skin envelope, while the selected orbital displays a right-skin envelope.

Figure~\ref{fig:HN} validates these statements for $N=40$, $t_R=1$, $t_L=0.17$, $\kappa=0.91$, $\Gamma=0.03$, $s_\ast=15$. The figure has three roles. Panel (a) compares the slow right-mode profile, the dominant natural orbital, and the normalized density. Panel (b) tests the source-position dependence predicted from the left eigenmode. Panel (c) shows the occupation hierarchy of natural orbitals. Together these checks verify that loading and geometry are controlled by different sides of the biorthogonal basis.

We next consider a nonreciprocal SSH chain. The hoppings are parameterized as $t_1^R=t_1\ee^g$, $t_1^L=t_1\ee^{-g}$, $ t_2^R=t_2\ee^g$, $t_2^L=t_2\ee^{-g}$, where $g$ controls the nonreciprocity. In the basis $(1A,1B,2A,2B,\ldots,NA,NB)$,
\begin{equation}
 X_{\rm SSH}=\kappa I-t_1^R T_1^R-t_1^L T_1^L-t_2^R T_2^R-t_2^L T_2^L,
 \label{eq:XSSH}
\end{equation}
where $T_1^R=\sum_n\ket{n,B}\bra{n,A}$, $T_1^L=(T_1^R)^\dagger$, $T_2^R=\sum_{n=1}^{N-1}\ket{n+1,A}\bra{n,B}$, and $T_2^L=(T_2^R)^\dagger$.
A diagonal similarity transformation maps $X_{\rm SSH}$ to a Hermitian SSH chain with couplings $t_1$ and $t_2$. In the topological regime $t_1<t_2$, the Hermitian reference problem has edge states. Transforming back gives right edge modes, left edge modes, and bulk modes with different nonreciprocal envelopes. The steady-state kernel can therefore select either a topological edge candidate or a slow bulk-skin candidate.

The numerical scan uses $N=20$, $t_1=0.50$, $t_2=1.00$, $\kappa=1.50$, $\Gamma=10^{-8}$, $(m,A)=(1,A)$. The weak pump keeps the source in the linear-probe regime. Since the plotted profiles and overlaps are normalized, their values are independent of the overall scale of $\Gamma$. We define
\begin{equation}
 \mathcal O_{\rm edge}=|\langle\widehat R_{\rm edge}|\varphi_{\max}\rangle|^2,
 \qquad
 \mathcal O_{\rm slow}=|\langle\widehat R_{\rm slow}|\varphi_{\max}\rangle|^2 .
 \label{eq:OedgeOslow}
\end{equation}
Here $\widehat R_{\rm edge}$ is the Euclidean-normalized right edge candidate and $\widehat R_{\rm slow}$ is the Euclidean-normalized slow bulk-skin candidate.

Figure~\ref{fig:SSH}(a) scans $g\in[-0.55,0.60]$. For negative $g$, the dominant natural orbital has large overlap with the edge candidate. For positive $g$, the overlap with the slow bulk-skin mode takes over. Representative points are $g_{\rm edge}=-0.25$, $ g_{\rm bulk}=0.20$. Panels (b) and (c) show the corresponding real-space profiles. The important point is not that the density must differ strongly from the selected orbital in every regime. Rather, the dominant natural orbital gives the clean mode-resolved identification of what the steady state has selected: an edge candidate in panel (b) and a slow bulk-skin candidate in panel (c).

\setcounter{secnumdepth}{0}  
\paragraph{Discussion.}
The main result of this Letter is the exact steady-state selection formula Eq.~\eqref{eq:Css_biorth}. It shows that the mixed steady state of a Gaussian open chain is organized by the same biorthogonal geometry that controls nonreciprocal relaxation, but not in a way that is exhausted by the density profile. Slow denominators, left-mode loading, and right-mode geometry enter separately. A local source selects components through the left eigenmodes, while the selected spatial profile is drawn by the right eigenmodes. The dominant natural orbital isolates that component.

The two examples demonstrate different levels of the same mechanism. The \HN{} chain verifies the source-selection law in the minimal bulk-skin geometry. The nonreciprocal SSH chain shows that natural-orbital locking can switch between different candidates when topology and nonreciprocity coexist. This provides a steady-state diagnostic that is mode-resolved, directly computable from $\Css$, and sharper than a density-only description.

\setcounter{secnumdepth}{0}  
\paragraph{Acknowledgments.}
We acknowledge the support of the National Natural Science Foundation of China (Grants No. 12275193, 11975166).

\setcounter{secnumdepth}{0}  
\paragraph{Data availability.}
The data that support the findings of this article are not publicly available. 
The data are available from the authors upon reasonable request.

\clearpage
\onecolumngrid
\appendix
\setcounter{section}{0}
\setcounter{equation}{0}
\setcounter{figure}{0}
\setcounter{table}{0}
\renewcommand{\theequation}{S\arabic{equation}}
\renewcommand{\thefigure}{S\arabic{figure}}
\renewcommand{\thetable}{S\arabic{table}}
\setcounter{secnumdepth}{2}

\begin{center}
{\large\bf Supplemental Material for ``Natural-orbital locking reveals hidden steady-state skin order in Gaussian open fermion chains''}\par\vspace{0.6em}
Y. T. Wang and X. Z. Zhang\par\vspace{0.2em}
College of Physics and Materials Science, Tianjin Normal University, Tianjin 300387, China
\end{center}
\vspace{0.8em}

This Supplemental Material provides the technical foundation for the results presented in the main text. We first derive the closed equation of motion for the one-body correlation matrix starting from a microscopic number-conserving quadratic Lindblad equation. We then solve the resulting Lyapunov equation and obtain the exact biorthogonal decomposition of the steady-state correlator. This derivation makes explicit how slow relaxation denominators, left-mode source loading, and right-mode spatial profiles combine to determine the dominant natural orbital of the nonequilibrium steady state. We next apply the general formula to the two target relaxation matrices used in the main text. For the Hatano--Nelson chain, we give the similarity transformation to a Hermitian reference chain and derive the analytic source-loading law. For the nonreciprocal SSH chain, we state the finite-size model, define the edge and bulk-skin candidates, and specify the overlap diagnostics used to identify the edge-lock to bulk-skin-lock crossover. Finally, we present an inverse-design construction showing how the target matrices can be embedded in microscopic Gaussian Lindblad dynamics by appropriate Hamiltonians and quantum jumps. The notation follows the main text throughout.

\section{Exact derivations and explicit model reductions}

We collect here the derivations that support the main text. The logic proceeds from the microscopic Lindblad equation to the closed one-body equation, then to the exact steady-state solution, and finally to the explicit model reductions used in the \HN{} and nonreciprocal SSH examples. The notation is kept identical to the main text.

Consider a number-conserving quadratic Hamiltonian
\begin{equation}
 H=\sum_{i,j=1}^{N} h_{ij} c_i^{\dagger} c_j,
 \qquad h=h^{\dagger},
\end{equation}
and linear loss and gain jumps,
\begin{equation}
 L^-_{\mu}=\sum_{j=1}^{N}(u_{\mu})_j c_j,
 \qquad
 L^+_{\nu}=\sum_{j=1}^{N}(v_{\nu})_j c_j^{\dagger}.
\end{equation}
The Lindblad equation is
\begin{align}
 \dot\rho={}&-\ii[H,\rho]
 +\sum_{\mu}\left(L^-_{\mu}\rho L^{ -\dagger}_{\mu}-\frac12\{L^{ -\dagger}_{\mu}L^-_{\mu},\rho\}\right)+\sum_{\nu}\left(L^+_{\nu}\rho L^{+\dagger}_{\nu}-\frac12\{L^{+\dagger}_{\nu}L^+_{\nu},\rho\}\right).
 \label{eq:Lindblad_SM_prb_rev}
\end{align}
We define the one-body correlator
\begin{equation}
 C_{ij}(t)=\Tr\!\left[\rho(t)c_j^{\dagger}c_i\right].
 \label{eq:Cdef_SM_prb_rev}
\end{equation}
The Hamiltonian part follows from
\begin{equation}
 [c_j^{\dagger}c_i,c_m^{\dagger}c_n]=\delta_{im}c_j^{\dagger}c_n-\delta_{nj}c_m^{\dagger}c_i,
\end{equation}
which gives
\begin{equation}
 \dot C\big|_{H}=-\ii[h,C].
 \label{eq:Hpart_SM_prb_rev}
\end{equation}
For the loss and gain sectors it is convenient to define the Gram matrices
\begin{equation}
 \Gamma^- = \sum_{\mu}\bm u_{\mu}\bm u_{\mu}^{\dagger},
 \qquad
 \Gamma^+ = \sum_{\nu}\bm v_{\nu}\bm v_{\nu}^{\dagger}.
 \label{eq:Gram_SM_prb_rev}
\end{equation}
A direct fermionic reduction gives
\begin{equation}
 \dot C\big|_-=-\frac12\{\Gamma^-,C\},
 \qquad
 \dot C\big|_+=-\frac12\{\Gamma^+,C\}+\Gamma^+.
 \label{eq:lossgain_SM_prb_rev}
\end{equation}
Combining Eqs.~\eqref{eq:Hpart_SM_prb_rev} and \eqref{eq:lossgain_SM_prb_rev}, we arrive at
\begin{equation}
 \dot C=-\ii[h,C]-\frac12\{\Gamma^-+\Gamma^+,C\}+\Gamma^+.
 \label{eq:closedCeq_SM_prb_rev}
\end{equation}
With
\begin{equation}
 X=\ii h+\frac{\Gamma^-+\Gamma^+}{2},
 \qquad
 Y=\Gamma^+,
\end{equation}
this becomes the compact equation quoted in the main text,
\begin{equation}
 \dot C=-XC-CX^{\dagger}+Y.
 \label{eq:compactCeq_SM_prb_rev}
\end{equation}

The exact steady-state solution follows by introducing
\begin{equation}
 \widetilde C(t)=\ee^{Xt}C(t)\ee^{X^{\dagger}t}.
\end{equation}
Differentiation gives
\begin{equation}
 \frac{d\widetilde C}{dt}=\ee^{Xt}Y\ee^{X^{\dagger}t},
\end{equation}
so that
\begin{equation}
 C(t)=\ee^{-Xt}C(0)\ee^{-X^{\dagger}t}+\int_0^t du\,\ee^{-Xu}Y\ee^{-X^{\dagger}u}.
 \label{eq:timeSol_SM_prb_rev}
\end{equation}
If all eigenvalues $\beta_n$ of $X$ satisfy $\Re\,\beta_n>0$, the first term decays and the steady state is
\begin{equation}
 \Css=\int_0^{\infty} dt\,\ee^{-Xt}Y\ee^{-X^{\dagger}t}.
 \label{eq:Cintegral_SM_prb_rev}
\end{equation}
Assume now that $X$ is diagonalizable,
\begin{equation}
 X\ket{R_n}=\beta_n\ket{R_n},
 \qquad
 X^{\dagger}\ket{L_n}=\beta_n^*\ket{L_n},
\end{equation}
with
\begin{equation}
 \langle L_m|R_n\rangle=\delta_{mn},
 \qquad
 \sum_n\ket{R_n}\bra{L_n}=I.
\end{equation}
Inserting the spectral decomposition of $\ee^{-Xt}$ into Eq.~\eqref{eq:Cintegral_SM_prb_rev} yields
\begin{equation}
 \Css=
 \sum_{m,n}
 \frac{\mel{L_m}{Y}{L_n}}{\beta_m+\beta_n^*}
 \ket{R_m}\bra{R_n}.
 \label{eq:Cssbiorth_SM_prb_rev}
\end{equation}
For a local pump $Y=\Gamma\ket{s}\bra{s}$ this simplifies to
\begin{equation}
 \Css=
 \Gamma\sum_{m,n}
 \frac{L_m^*(s)L_n(s)}{\beta_m+\beta_n^*}
 \ket{R_m}\bra{R_n}.
 \label{eq:Csslocal_SM_prb_rev}
\end{equation}
If one mode is parametrically slower than the rest, then
\begin{equation}
 \Css\approx \mathcal A_0(s)\ket{R_0}\bra{R_0},
 \qquad
 \mathcal A_0(s)=\frac{\Gamma|L_0(s)|^2}{2\Re\,\beta_0}.
\end{equation}
Because $\ket{R_0}$ is biorthogonally normalized rather than Euclidean normalized, the dominant natural orbital is approximately
\begin{equation}
 \ket{\varmax}\approx \ket{\widehat R_0}
 =\frac{\ket{R_0}}{\sqrt{\langle R_0|R_0\rangle}},
 \qquad
 \numax\approx \mathcal A_0(s)\langle R_0|R_0\rangle.
\end{equation}

The numerical diagnostics used in the main text are defined as follows. The normalized density is
\begin{equation}
 \trn_j = \frac{n_j^{\mathrm{ss}}}{\sum_{\ell}n_{\ell}^{\mathrm{ss}}},
 \qquad n_j^{\mathrm{ss}}=(\Css)_{jj}.
 \label{eq:trn_SM_prb_rev}
\end{equation}
For the Hatano--Nelson validation in Fig.~2(c) of the main text, the normalized occupation spectrum is
\begin{equation}
 \trnu_\alpha=\frac{\nu_\alpha}{\numax},
 \qquad \numax=\nu_1.
 \label{eq:trnu_SM_prb_rev}
\end{equation}
The Euclidean-normalized right eigenmode is
\begin{equation}
 \ket{\widehat R_n}=\frac{\ket{R_n}}{\sqrt{\langle R_n|R_n\rangle}},
\end{equation}
and its overlap with the dominant natural orbital is
\begin{equation}
 \mathcal O_n=\left|\langle \widehat R_n|\varmax\rangle\right|^2.
 \label{eq:On_SM_prb_rev}
\end{equation}
Whenever we compare mode-resolved source loadings, we use
\begin{equation}
 \mathcal A_n(s)=\frac{\Gamma |L_n(s)|^2}{2\Re\,\beta_n},
 \qquad
 \trA_n = \frac{\mathcal A_n}{\max_m \mathcal A_m}.
 \label{eq:An_SM_prb_rev}
\end{equation}
In the SSH scan we define the edge candidate as the mode whose eigenvalue is closest to $\kappa$ and whose right-state weight is largest near the appropriate boundary. The bulk candidate is the slowest mode $n_{\mathrm{slow}}=\arg\min_n \Re\,\beta_n$.

We next summarize the model reductions used in the main text. For the nonreciprocal \HN{} chain,
\begin{equation}
 X_{\mathrm{HN}}=\kappa I-t_R\sum_{j=1}^{N-1}\ket{j+1}\bra{j}-t_L\sum_{j=1}^{N-1}\ket{j}\bra{j+1},
\end{equation}
define
\begin{equation}
 r=\sqrt{\frac{t_R}{t_L}},
 \qquad
 S=\sum_{j=1}^{N}r^j\ket{j}\bra{j}.
\end{equation}
Then
\begin{equation}
 S^{-1}X_{\mathrm{HN}}S=
 \kappa I-\sqrt{t_Rt_L}\sum_{j=1}^{N-1}\left(\ket{j+1}\bra{j}+\ket{j}\bra{j+1}\right).
\end{equation}
The Hermitian reference eigenfunctions are
\begin{equation}
 \phi_n(j)=\sqrt{\frac{2}{N+1}}\sin\frac{n\pi j}{N+1},
\end{equation}
with eigenvalues
\begin{equation}
 \beta_n=\kappa-2\sqrt{t_Rt_L}\cos\frac{n\pi}{N+1}.
\end{equation}
Transforming back gives
\begin{equation}
 R_n(j)=r^j\phi_n(j),
 \qquad
 L_n(j)=r^{-j}\phi_n(j).
 \label{eq:HNRL_SM_prb_rev}
\end{equation}
For a local pump $Y=\Gamma\ket{s}\bra{s}$,
\begin{equation}
 C_{jk}^{\mathrm{ss}}=
 \Gamma r^{j+k-2s}
 \sum_{m,n}
 \frac{\phi_m(s)\phi_n(s)\phi_m(j)\phi_n(k)}{\beta_m+\beta_n}.
 \label{eq:HNkernel_SM_prb_rev}
\end{equation}
In the single-slow-mode regime, the source-dependent loading factor is
\begin{equation}
 \mathcal A_1(s)=\frac{\Gamma r^{-2s}}{2\beta_1}\frac{2}{N+1}\sin^2\frac{\pi s}{N+1}.
\end{equation}
The dominant natural orbital is the Euclidean-normalized version of the right mode,
\begin{equation}
 \varmax(j)\approx \frac{r^j\sqrt{\frac{2}{N+1}}\sin\frac{\pi j}{N+1}}
 {\left[\sum_{\ell=1}^{N}r^{2\ell}\frac{2}{N+1}\sin^2\frac{\pi \ell}{N+1}\right]^{1/2}},
 \qquad
 \numax\approx \mathcal A_1(s)\langle R_1|R_1\rangle.
 \label{eq:HNsourcelaw_SM_prb_rev}
\end{equation}

For the nonreciprocal SSH chain we use the parameterization $t_1^R=t_1\ee^g$, $t_1^L=t_1\ee^{-g}$, $ t_2^R=t_2\ee^g$, $t_2^L=t_2\ee^{-g}$. The target matrix is
\begin{align}
 X_{\mathrm{SSH}}={}&\kappa I
 -\sum_{n=1}^{N}\left(t_1^R\ket{n,B}\bra{n,A}+t_1^L\ket{n,A}\bra{n,B}\right)
 \notag\\
 &-\sum_{n=1}^{N-1}\left(t_2^R\ket{n+1,A}\bra{n,B}+t_2^L\ket{n,B}\bra{n+1,A}\right).
 \label{eq:XSSH_SM_prb_rev}
\end{align}
A diagonal similarity transformation reduces $X_{\mathrm{SSH}}$ to a Hermitian SSH chain with effective couplings $t_1$ and $t_2$. In the topological regime $t_1<t_2$, the Hermitian reference chain supports edge states. For the left edge mode one has
\begin{equation}
 \phi_{\mathrm{edge},L}(n,A)\propto\left(-\frac{t_1}{t_2}\right)^{n-1},
 \qquad
 \phi_{\mathrm{edge},L}(n,B)\approx 0.
\end{equation}
Transforming back gives the right and left edge envelopes
\begin{equation}
 R_{\mathrm{edge}}(n,A)\propto\left(-\frac{t_1^R}{t_2^L}\right)^{n-1},
 \qquad
 L_{\mathrm{edge}}(n,A)\propto\left(-\frac{t_1^L}{t_2^R}\right)^{n-1}.
 \label{eq:SSHedge_SM_prb_rev}
\end{equation}
For a local pump on the $A$ site of cell $m$,
\begin{equation}
 Y=\Gamma\ket{m,A}\bra{m,A},
\end{equation}
the exact loading spectrum follows from Eq.~\eqref{eq:An_SM_prb_rev}. In the numerics used in the main text we keep a common system size and source geometry, the common parameters are $N=20$, $t_1=0.50$, $t_2=1.00$, $\kappa=1.50$, $\Gamma=10^{-8}$, $(m,A)=(1,A)$. Within this common setup, the edge-locking representative point is $g_{\mathrm{edge}}=-0.25$, and the bulk-skin-locking representative point is $g_{\mathrm{bulk}}=0.20$.

In the crossover scan $ g\in[-0.55,0.60]$, we compare the two overlaps
\begin{equation}
 \mathcal O_{\mathrm{edge}}=|\langle \widehat R_{\mathrm{edge}}|\varphi_{\mathrm{max}}\rangle|^2,
 \qquad
 \mathcal O_{\mathrm{slow}}=|\langle \widehat R_{\mathrm{slow}}|\varphi_{\mathrm{max}}\rangle|^2.
 \label{eq:OedgeOslow_SM_prb_rev}
\end{equation}
The edge overlap dominates for negative $g$, while the slow bulk-skin overlap takes over for sufficiently positive $g$. This is the switching diagnostic used in Fig.~3(a) of the main text.

\section{Inverse-design construction of the microscopic open systems}

We now summarize the inverse-design problem that reconstructs a microscopic quadratic open system from a target pair $(X_{\mathrm{tar}},Y_{\mathrm{tar}})$. The general relation is
\begin{equation}
 X=\ii h+\frac{\Gamma^-+\Gamma^+}{2},
 \qquad
 Y=\Gamma^+.
 \label{eq:inverse_general_SM_prb_rev}
\end{equation}
Hence the formal solution is
\begin{equation}
 h=\frac{X_{\mathrm{tar}}-X_{\mathrm{tar}}^{\dagger}}{2\ii},
 \qquad
 \Gamma^+=Y_{\mathrm{tar}},
 \qquad
 \Gamma^-=X_{\mathrm{tar}}+X_{\mathrm{tar}}^{\dagger}-Y_{\mathrm{tar}}.
 \label{eq:inverse_solution_SM_prb_rev}
\end{equation}
The task is then to realize the positive semidefinite matrices $\Gamma^+$ and $\Gamma^-$ through explicit local quantum jumps. In the most general diagonal-pump implementation one may take
\begin{equation}
 Y_{\mathrm{tar}}=\sum_j y_j\ket{j}\bra{j},
 \qquad L_j^+=\sqrt{y_j}\,c_j^\dagger,
\end{equation}
with $y_j\ge0$. A local pump is the special case $y_j=\Gamma\delta_{j,s}$. The residual onsite loss strengths in the decompositions below are then obtained by replacing the uniform pump strength $\gamma$ by the corresponding site-dependent $y_j$. The simple uniform-pump formulas are written explicitly because they are compact and show the structure of the construction.

For the \HN{} target matrix,
\begin{equation}
 X_{\mathrm{HN}}=\kappa I-t_R T-t_L T^{\dagger},
 \qquad
 T=\sum_{j=1}^{N-1}\ket{j+1}\bra{j},
\end{equation}
we choose a uniform pump,
\begin{equation}
 Y_{\mathrm{tar}}=\gamma I,
 \qquad
 L_j^+=\sqrt{\gamma}\,c_j^{\dagger},
 \qquad j=1,\dots,N.
\end{equation}
The Hamiltonian follows from Eq.~\eqref{eq:inverse_solution_SM_prb_rev},
\begin{equation}
 H=\frac{\ii}{2}(t_R-t_L)\sum_{j=1}^{N-1}\left(c_{j+1}^{\dagger}c_j-c_j^{\dagger}c_{j+1}\right).
\end{equation}
The remaining loss matrix is
\begin{equation}
 \Gamma^-=(2\kappa-\gamma)I-(t_R+t_L)(T+T^{\dagger}).
\end{equation}
A transparent local decomposition is obtained by introducing the bond jumps
\begin{equation}
 B_j^-=\sqrt{t_R+t_L}\,(c_j-c_{j+1}),
 \qquad j=1,\dots,N-1,
\end{equation}
and the onsite losses
\begin{equation}
 S_1^-=\sqrt{\delta-\beta}\,c_1,
 \qquad
 S_N^-=\sqrt{\delta-\beta}\,c_N,
\end{equation}
\begin{equation}
 S_j^-=\sqrt{\delta-2\beta}\,c_j,
 \qquad j=2,\dots,N-1,
\end{equation}
with
\begin{equation}
 \delta=2\kappa-\gamma,
 \qquad
 \beta=t_R+t_L.
\end{equation}
This explicit construction is valid under the simple sufficient condition
\begin{equation}
 2\kappa-\gamma \ge 2(t_R+t_L).
\end{equation}

For the nonreciprocal SSH target matrix,
\begin{align}
 X_{\mathrm{SSH}}={}&\kappa I
 -\sum_{n=1}^{N}\left(t_1^R\ket{n,B}\bra{n,A}+t_1^L\ket{n,A}\bra{n,B}\right)-\sum_{n=1}^{N-1}\left(t_2^R\ket{n+1,A}\bra{n,B}+t_2^L\ket{n,B}\bra{n+1,A}\right),
\end{align}
we again choose a uniform pump,
\begin{equation}
 Y_{\mathrm{tar}}=\gamma I,
 \qquad
 L_{n,A}^+=\sqrt{\gamma}\,c_{n,A}^{\dagger},
 \qquad
 L_{n,B}^+=\sqrt{\gamma}\,c_{n,B}^{\dagger}.
\end{equation}
The Hamiltonian is
\begin{align}
 H={}&\frac{\ii}{2}(t_1^R-t_1^L)\sum_{n=1}^{N}\left(c_{n,B}^{\dagger}c_{n,A}-c_{n,A}^{\dagger}c_{n,B}\right)+\frac{\ii}{2}(t_2^R-t_2^L)\sum_{n=1}^{N-1}\left(c_{n+1,A}^{\dagger}c_{n,B}-c_{n,B}^{\dagger}c_{n+1,A}\right).
\end{align}
The Hermitian part that must be reproduced dissipatively is
\begin{align}
 \Gamma^-={}&(2\kappa-\gamma)I
 -(t_1^R+t_1^L)\sum_{n=1}^{N}\left(\ket{n,B}\bra{n,A}+\ket{n,A}\bra{n,B}\right)-(t_2^R+t_2^L)\sum_{n=1}^{N-1}\left(\ket{n+1,A}\bra{n,B}+\ket{n,B}\bra{n+1,A}\right).
\end{align}
A local decomposition is obtained from the intracell bond losses
\begin{equation}
 B_{1,n}^-=\sqrt{t_1^R+t_1^L}\,(c_{n,A}-c_{n,B}),
 \qquad n=1,\dots,N,
\end{equation}
and the intercell bond losses
\begin{equation}
 B_{2,n}^-=\sqrt{t_2^R+t_2^L}\,(c_{n,B}-c_{n+1,A}),
 \qquad n=1,\dots,N-1.
\end{equation}
These bond jumps generate the desired nonzero off-diagonal matrix elements and part of the diagonal ones. The remaining diagonal weights are supplied by onsite losses. Writing
\begin{equation}
 \delta=2\kappa-\gamma,
 \qquad
 \beta_1=t_1^R+t_1^L,
 \qquad
 \beta_2=t_2^R+t_2^L,
\end{equation}
the required onsite terms are
\begin{equation}
 S_{1,A}^-=\sqrt{\delta-\beta_1}\,c_{1,A},
 \qquad
 S_{N,B}^-=\sqrt{\delta-\beta_1}\,c_{N,B},
\end{equation}
\begin{equation}
 S_{n,A}^-=\sqrt{\delta-\beta_1-\beta_2}\,c_{n,A},
 \qquad n=2,\dots,N,
\end{equation}
\begin{equation}
 S_{n,B}^-=\sqrt{\delta-\beta_1-\beta_2}\,c_{n,B},
 \qquad n=1,\dots,N-1.
\end{equation}
A simple sufficient condition for this explicit local realization is
\begin{equation}
 2\kappa-\gamma \ge (t_1^R+t_1^L)+(t_2^R+t_2^L).
\end{equation}
The same construction applies to a diagonal local pump by using the site-dependent replacement described above. A sufficiently large uniform damping shift $\kappa$ guarantees positivity of the residual loss matrix while preserving the nonreciprocal eigenvector geometry of the target family.

\end{document}